\documentclass[12pt,a4paper]{article}

\usepackage[utf8]{inputenc}
\usepackage{amsmath}
\usepackage{amssymb}
\usepackage[toc,title]{appendix}
\usepackage{graphicx}
\usepackage{color}
\usepackage{hyperref}
\usepackage{authblk}
\usepackage{cite}
\usepackage{mcite}

\DeclareMathOperator*{\tr}{tr}

\newcommand{\bra}[1]{\langle #1 |}
\newcommand{\ket}[1]{| #1 \rangle}  
 
\begin{document} 
 
\title{Scrambling in the Black Hole Portrait}
\date{July 12, 2013}

\author[a,b,c]{Gia Dvali}

\author[a]{Daniel Flassig\thanks{daniel.flassig@physik.lmu.de}}

\author[a,d]{Cesar Gomez\thanks{cesar.gomez@uam.es}}

\author[a]{\\Alexander Pritzel\thanks{alexander.pritzel@physik.lmu.de}}

\author[a]{Nico Wintergerst\thanks{nico.wintergerst@physik.lmu.de}}

\affil[a]{Arnold Sommerfeld Center for Theoretical Physics, Fakult\"at f\"ur
Physik, Ludwig-Maximilians-Universit\"at M\"unchen, Theresienstr.~37, 80333 M\"unchen,
Germany}

\affil[b]{Max-Planck-Institut f\"ur Physik\\
F\"ohringer Ring 6, 80805 M\"unchen, Germany}

\affil[c]{Center for Cosmology and Particle Physics\\
Department of Physics, New York University\\
4 Washington Place, New York, NY 10003, USA}

\affil[d]{Instituto de F\'{\i}sica Te\'orica UAM-CSIC, C-XVI \\
Universidad Aut\'onoma de Madrid,
Cantoblanco, 28049 Madrid, Spain}

\maketitle

\begin{abstract}
  Recently a quantum portrait of black holes was suggested according to which a macroscopic  black hole  is a Bose-Einstein condensate of soft gravitons stuck at the critical point of a quantum phase transition.    We explain why quantum criticality and instability are  the key for efficient generation of entanglement and consequently of the scrambling of information.   By studying a simple Bose-Einstein prototype, we 
   show that the scrambling time, which is set by the quantum break time of the system, goes as $\log  N \,$  for $N$ the number of quantum constituents or equivalently the black hole entropy.   
\end{abstract} 
\newpage
\section{Introduction}
The present state of affairs in black hole physics is somewhat paradoxical. On one side, it is widely believed that the final state of the black hole evaporation process is a pure state,  while on the other side, the standard Hawking's model of evaporation does not account for the purification mechanism. Obviously the missing ingredient is a microscopic quantum model of the black hole beyond its pure geometrical definition. 

 In the present paper, we shall focus on a specific microscopic description, put forward  in 
 \cite{Dvali:2011aa, Dvali:2012en, Dvali:2012wq,Dvali:2012rt, Flassig:2012re} (for different aspects of this portrait, see \cite{berkhahn2013microscopic,*Mueck:2013mha,*Binetruy:2012kx,*Casadio:2013hja,*Dvali:2012uq}). In this picture, black holes of arbitrarily large size $R$ are treated as self-sustained bound states 
 of a large number of long wavelength ($\sim R$) gravitons.   From the quantum physics point of view, such a  
 bound-state represents  a Bose-Einstein condensate stuck at the critical point of  a quantum phase transition. 
 This quantum criticality is the key to the understanding of the  mysterious properties of black holes
 that emerge in the naive semi-classical treatment.
    
 In this respect, our approach sharply differs from previous attempts, such as 
 $D$-brane models for extremal black holes\cite{strominger1996microscopic}, models based on Matrix theory \cite{Banks:1996vh,*Banks:1997hz,*Banks:1997tn} and fuzzballs \cite{mathur2005fuzzball}. 
   These approaches heavily rely on a particular UV-completion of gravity at short distances, such as string or Planck ($L_p$) length-scales.  Our key postulate is fundamentally different. 
  We state that physics of macroscopic  black holes of size $R \, \gg \, L_P$,  must be largely insensitive to the properties of UV-completion 
 at Planck-distances and must be governed solely by the quantum physics of long wavelength gravitons  with their quantum interaction strength being fully determined by the graviton-graviton 
 interaction vertices of Einstein theory.  All the seemingly-mysterious properties of 
 the black holes must originate  from  {\it collective quantum}  phenomena of these constituent soft gravitons.   To put it shortly,  in our picture large black holes are not governed by UV-physics, but rather 
 by the quantum collective effects of IR-physics. 
 
     These collective effects render the entire macroscopic system extremely sensitive to quantum effects.  A fundamental aspect is the 
  appearance of large number of almost gapless collective modes (Bogolyubov modes), which
  can be thought of as the quantum holographic degrees of freedom. 
   They are responsible for the instability of the condensate, for its quantum depletion as well as for a large (near)degeneracy of the quantum states.    
    These phenomena provide the underlying quantum-mechanical dynamics for black hole evaporation, entropy and holography.  
   
 An accompanying property of the quantum phase transition is a very efficient generation of entanglement. Sharp rise of one-particle ground-state entanglement was already confirmed by numerical studies of a prototype model \cite{Flassig:2012re}.   
 
 In this paper, we shall discuss how the instability of the BEC is the key for understanding the 
 efficient generation of entanglement and information-scrambling by a black hole in a 
 logarithmic time, 
 \begin{equation}
   t_\mathrm{scrambling}/R \, \propto \, \log  N \, .  
  \label{scrambling}
  \end{equation} 
  Noticing that in our treatment $N$ measures the number of constituents, this result is in 
 full agreement with the semi-classical prediction originally made in \cite{Hayden:2007cs, Sekino:2008he}. 
  
 Let us briefly review some of the key ingredients of the black hole quantum portrait. 
In the picture of \cite{Dvali:2012en}  we track the formation of a black hole as bringing the graviton condensate to  {\it the critical point of a quantum phase transition}. 
 At this point the BE condensate is nearly self-sustained with mass $M$ and size $R$ related to the total number $N$ of constituents as $M = \sqrt{N} L_P^{-1},  \, R = \sqrt{N} L_P$.  However, 
 the condensate is unstable both with respect to 
collapse as well as to quantum depletion. 
  The two effects balance each other in such a way that although the condensate slowly collapses and loses its gravitons, it stays at the quantum critical point.  This process can be parametrized as 
 a self-similar decrease of $N$,   
  \begin{equation}
   \frac{dN}{dt} \, = \, -\frac {1}{\sqrt{N} L_P} \, .  
   \label{decrease}
   \end{equation}
Note that this instability survives in the semi-classical limit ($L_P \to 0, N \to \infty,   \sqrt{N} L_P =$ fixed), which 
corresponds to the Gross-Pitaevskii limit of the graviton condensate.  

One of the most important outputs of the black hole $N$-quantum portrait is to allow us to identify important quantum corrections that are not resolvable within the standard semi-classical approximation.  In the semi-classical picture one works with the notion of classical metric. Irrespectively
whether the metric is derived from the loop-corrected effective action, it is an intrinsically classical entity and its quantum constituents are  
not resolved. The only non-perturbative quantum corrections that one can visualise in this limit for a black hole of action $S$ are of the form $e^{-\frac{S}{\hbar}}$.  These sort of corrections take into the account only the total black hole action and are blind to any form of microscopic constituency. 
Such corrections, for instance, can measure the transition amplitudes between black hole and thermal topologies \cite{Maldacena:2001kr,Hawking:2005kf}. 

 On the other hand there exist more important quantum corrections that scale as
$\hbar/S$, but they are unaccountable in the semi-classical treatment. The key problem  lies in unveiling their  {\it microscopic meaning} as well as in  understanding under what conditions these quantum corrections can effectively lead to order-one effects  for macroscopic black holes.
In the quantum $N$-portrait these corrections  naturally appear as 
$1/N$ corrections, since the occupation number of gravitons measures the black hole action (as well as the entropy),    
\begin{equation}
  N = \frac{S}{\hbar} \, . 
\label{NandS}
\end{equation}
 Thus, the quantity $1/N$ is a measure of quantum effects that are much more important then the $e^{-N}$-type effects captured by the 
semi-classical analysis. 
 In particular, it was shown that $1/N$-corrections account for the deviations from thermality of black hole radiation \cite{Dvali:2011aa} as well as  
for the quantum hair of black holes \cite{Dvali:2012rt}.  Existence of these corrections was also confirmed  for the string holes \cite{Veneziano:2012yj}.\footnote{The similarly large corrections are also indicated  in a different treatment in which one prescribes a wave-function to the horizon \cite{Brustein:2013qma,*Brustein:2013xga,*Casadio:2013aua}, This approach differs from ours since the metric is still treated semi-classically and its quantum constituents are not resolved. Nevertheless the largeness  of the corrections is in a qualitative agreement.}  These  $1/N$-corrections are the key for abolishing the black hole "information paradox",  since
over the black hole half-lifetime  they give order-one effect for arbitrarily-large black holes $N \gg 1$ \cite{Dvali:2012wq}.

%One of the most important outputs of the black hole portrait is to allow us to organize quantum corrections in a natural way. For a black hole of action (or entropy)  $S \, = \, \hbar \, N$ we have non-perturbative corrections that go as $e^{-N}$. These sort of corrections take into the account only the total black hole action and are blind to any form of microscopic constituency. They can for instance measure the transition amplitude between black hole and thermal topologies \cite{Maldacena:2001kr,Hawking:2005kf}. On the other hand, any form of quantum back-reaction should be measured in terms of the ratio $1/N \sim \frac{\hbar}{S}$. The key problem however lies in unveiling the {\it microscopic meaning} of $1/N$-corrections as well as in  understanding under what conditions these quantum corrections can effectively become order one for macroscopic black holes with $N\, \gg \, 1$. 

A Bose-Einstein  condensate represents a very natural setup for identifying the physical meaning of $1/N$-corrections. In a nutshell, for BE condensates the small quantum deviations from the mean field Gross-Pitaevskii (GP) description are 
$1/N$-corrections, with $1/N$ replacing the role of the Planck constant $\hbar$. Moreover, as we will discuss in this paper, instabilities of the GP equation can naturally lead to {\it fast} enhancement of these quantum corrections. More concretely, around instabilities of the GP equation the {\it quantum break time}  (i.e.  the time needed to depart significantly ($O(1)$) from the mean field approximation) scales with $N$ as $\log N$. Nicely enough, the BE portrait of black holes implies instabilities of the GP equation. The root of these instabilities lies in the mean-field instability of the condensate  at the quantum critical point due to the attractive nature of the interaction.  As we will show in this note, the quantum break time for BE condensates fits naturally with the notion of scrambling time for black holes.

\section{Scrambling and Quantum Break Time}
The notion of black holes as scramblers was first introduced in \cite{Hayden:2007cs}, where it was realized that perturbed black holes should thermalize in a time $t \geq R \log S_{BH}$ for $S_{BH}$ the black hole entropy and $R$ the black hole radius. In \cite{Sekino:2008he} it was then suggested that black holes may saturate this bound, a property that has become known as fast scrambling. The associated timescale is now known as scrambling time.\footnote{For several attempts to understand the physics of scrambling, see \cite {Susskind2,Lashkari:2011yi,Asplund:2011qj,Barbon:2011pn}.}

The concept of scrambling is intimately related to entanglement of subsystems. Consider a quantum mechanical system whose Hilbert space is a direct product $\mathcal H = \mathcal H_A \otimes \mathcal H_B$ in a state described by the density matrix $\rho$. The conventional measure of entanglement between the subsystems is the Von Neumann entropy of the reduced density matrix:
\begin{equation}
	S_A = \tr_A \left( \rho_A \log \rho_A \right)
	\quad
	\rho_A = \tr_B \rho
\end{equation}
A system is called a scrambler if it dynamically thermalizes in the sense that, if prepared in an atypical state, it evolves towards typicality. That is, even for an initial state that has little or no entanglement between subsystems, the time evolution is such that the reduced density matrices are finally close to thermal density matrices. 
The scrambling time is simply the characteristic time scale associated to this process. It can be described as the time it takes for a perturbed system, one that is described by a product state, to evolve back into a strongly entangled state. It can also be interpreted as the time necessary to distribute any information entering the system amongst  all its constituents.

The quantum meaning of the scrambling time becomes more transparent if we rewrite it as
\begin{equation}
t_\mathrm{scrambling}\sim R \log \left( \frac{S}{\hbar} \right)
\end{equation}
with $S$ now denoting the action of the black hole. This is the typical expression for the \emph{quantum break time} provided the system is near an instability, where quantum break time denotes the timescale for the breakdown of the classical (mean field) description. Hence we will identify as a necessary condition for a system to behave as a fast scrambler to have a quantum break time scaling logarithmically with the number of constituents.

\subsection{Logarithmic Quantum Break Time}
In the context of quantum chaos, it has long been known that under certain conditions, the classical description breaks down much quicker than the naively expected polynomial quantum break time. Specifically, in the vicinity of an instability for the classical description, i.e. positive local Lyapunov exponent $\lambda$, the quantum break time usually goes as 
\begin{equation}
	t_\mathrm{break} \sim \lambda^{-1} \log \frac{S}{\hbar}
\end{equation}
This exactly resembles the logarithmic scaling of the scrambling time. In fact, the black hole scrambling time coincides with the typical quantum break time if the microscopic description of the black hole contains an instability characterized by a Lyapunov exponent $\lambda \sim 1/R$. The black hole quantum portrait contains such an instability which survives in the semi-classical limit ($L_P=0$, $N=\infty$,  with $\sqrt{N}L_P$ fixed ) and is described by equation (\ref{decrease}). The characteristic timescale is given by $R=\sqrt{N}L_P$ which classically becomes the black hole radius. Hence we expect the Lyapunov exponent to be set by $1/R$.
 This is precisely the way we will identify scrambling in the BE portrait of black holes. 

For the convenience of the interested reader, in appendix \ref{husimi-argument} we reproduce a general argument for logarithmic quantum break time at an instability. In the next section we show specifically for Bose-Einstein condensate systems that they exhibit quantum breaking in the scrambling time. We will also comment on the instability there. In section \ref{numerics}, we perform a numerical analysis that confirms this reasoning.
 
\subsection{Chaos and Thermalization}
The relation between scrambling and quantum break time is even stronger if the classical limit of the relevant system not only contains a local instability, but also exhibits classical chaos. For such systems it has been claimed - and checked to some extent - that the time scale of thermalization is of the same order as $t_\mathrm{break}$ \cite{altland2012quantum}. By taking a pure quantum state it was shown that the time evolution not only stretches and folds the quasi-probability distribution, but also smoothens it out. Of course the quantum state stays pure, but it is thermalized in the sense of being smeared out over the accessible classical phase space volume. This would presumably imply scrambling as defined above. Although, at this point we cannot prove that this is indeed how scrambling actually takes place in the graviton condensates of the BH portrait,  we do take it as further evidence that the quantum break time is intimately related with scrambling time. 

\section{Quantum Break Time in BE Condensates}

\subsection{Prototype Models}
It has been pointed out \cite{Dvali:2011aa,Dvali:2012en,Dvali:2012wq} that many of the seemingly mysterious properties of black holes can be resolved when considering them as Bose-Einstein  condensates of long wavelength gravitons that interact with a critical coupling strength. Indeed, it has been realized that a vast amount of those properties can already be explored in much simpler systems. These systems share the crucial property that they contain bifurcation or quantum critical points. 

Within this work we will follow that route and further explore models of attractive cold bosons both in one and three spatial dimensions. We will show that they exhibit a logarithmic quantum break time, again intimately related to the existence of instabilities and quantum critical or bifurcation points. 

The explicit models under consideration in d+1 dimensions are described by the Hamiltonian
\begin{equation}
	H = \int_V d^dx \left(\frac{\hbar^2}{2m}(\nabla \phi^\dagger)(\nabla \phi) - \frac{g}{2} (\phi^\dagger \phi)^2\right) .
	\label{eq:hamiltonian}
\end{equation}
Here, $\phi$ carries the dimension $\mathrm{length}^{-d/2}$, while the coupling constant $g$ carries dimension $\mathrm{energy} \times \mathrm{length}^d$.
The integral is taken over the volume of a $d$-dimensional torus $V$.

Expanding $\phi$ into mean field and quantum fluctuations $\phi = \phi_\mathrm{mf} + \delta\phi$ and subsequent minimization of the energy functional
leads, at zeroth order, to the Gross-Pitaevskii (GP) equation for stationary solutions:
\begin{equation}
	i\hbar\partial_t \phi_\mathrm{mf} =\left(\frac{\hbar^2}{2m}\Delta + g |\phi_\mathrm{mf}|^2\right)\phi_\mathrm{mf} = \mu \phi_\mathrm{mf} .
	\label{eq:gross-pitaevskii}
\end{equation}
The chemical potential $\mu$ appears as a Lagrange multiplier that imposes a constraint on the particle number $N$, $\int_V d^dx \phi^\dagger \phi \, = \, N$.

\begin{figure}[t]
  \centering
  \includegraphics[width=\linewidth,angle=0.]{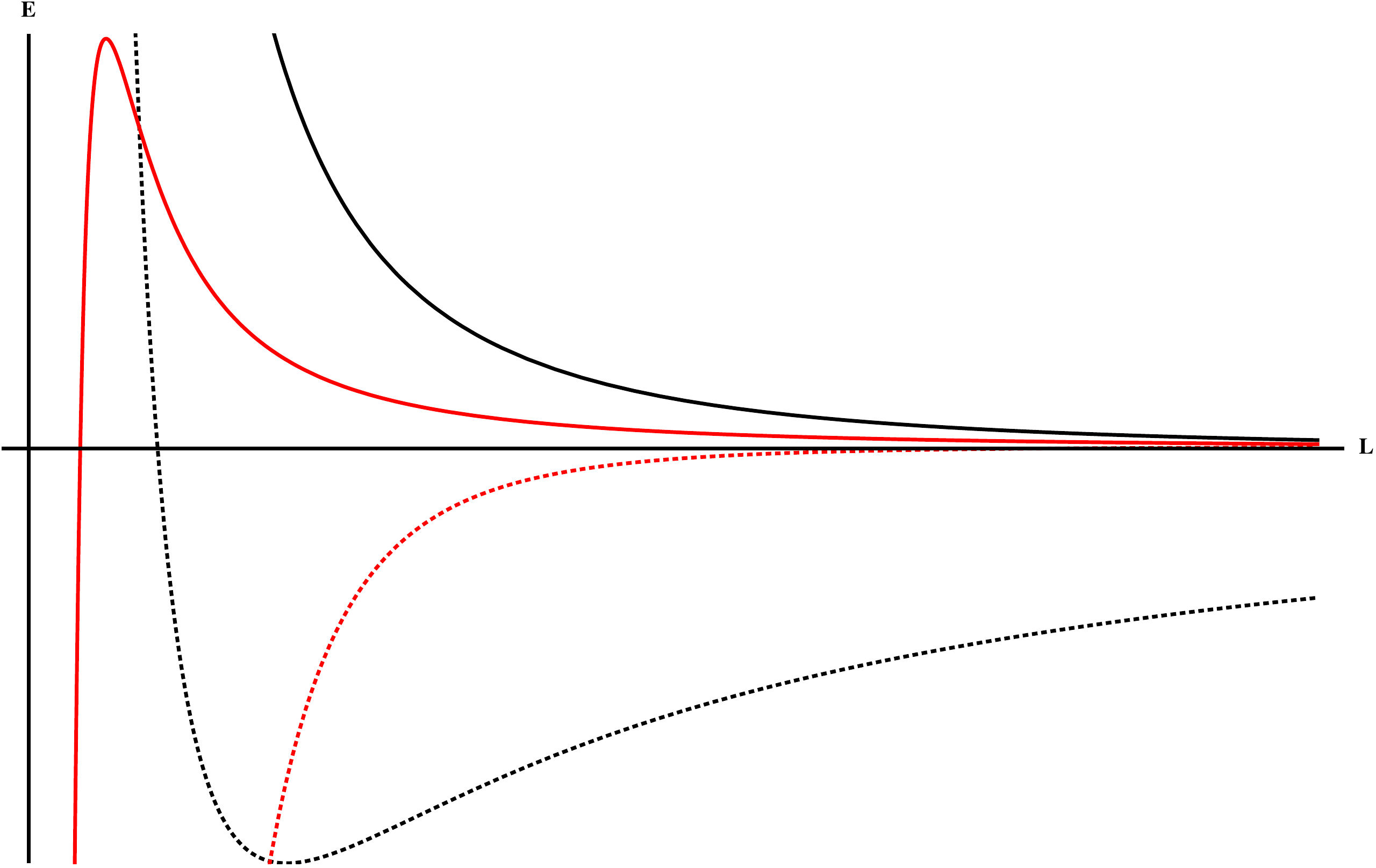}
  \caption{Energy as a function of the condensate width for $gN \ll 1$ (solid) and $gN \gg 1$ (dashed) for a condensate in 1 (black) and 3 (red) spatial dimensions.}
    \label{fig:escal}
\end{figure}

An intuitive understanding of the physics of these Bose-Einstein  condensates may be gained by considering the behavior of the energy when rescaling the characteristic size of the condensate $R$:
\begin{equation}
	E \sim \frac{N}{R^{2}} - gN\frac{N}{R^{d}}\,,
	\label{eq:escal}
\end{equation} 
where the coefficients of both terms naturally depend on the shape of the condensate. As illustrated in Fig. \ref{fig:escal}, the behavior depends strongly on the dimension under consideration. For $d=1$, the energy is always bounded from below. The (stable) ground state solution is given by a homogeneous condensate for $gN < 1$ and a localized soliton for $gN > 1$. A quantum phase transition is observed \cite{Kanamoto:2002xy} at $gN=1$. On the other hand, for $d \geq 3$, there is a classically stable homogeneous solution for $gN < 1$, while the condensate is unstable for $gN > 1$.

\subsection{Quantum Breaking in Bose Condensates}
We will now apply the notion of quantum breaking to a Bose-Einstein  condensate system of $N$ identical particles. In general, we want to study $k$-particle subsets (although $k$ particles do not form a proper subspace, this technicality will not disturb us much) and use the conventional $k$-particle sub-density matrices
\begin{equation}
	\rho^{(k)}_{m n} = \mathcal{N} \: \tr\left[\rho \left(\prod_l (a_l^\dagger)^{m_l}\right) \left(\prod_l a_l^{n_l}\right)\right]
\end{equation}
where $m$ and $n$ label $k$-particle states, $a_l$ is the annihilation operator for one Boson in the $l$ orbital and $n_l$ is the occupation number in state $n$ of orbital $l$, which satisfy $\sum n_l = k$. The normalization $\mathcal{N}$ is chosen so that $\tr \rho^{(1)} = 1$. We would identify a Bose gas as a fast scrambler, {\it if its time evolution would create a large entropy in each $\rho_k$ for $k \ll N$ on a timescale that scales logarithmically with $N$}.

More precisely, we do not expect generic atomic Bose-Einstein  condensates - as available in the laboratory - to be scramblers in the sense of the previous paragraph.  We do however identify the scrambling timescale to be the relevant thermalization scale for the quantum time 
evolution of the system in the following restricted way. 
If we do not insist on the thermalization of all sub-density matrices, but restrict our attention to $\rho^{(1)}$, then the time in which a state with pure $\rho^{(1)}$ develops a large von Neumann entropy in $\rho^{(1)}$ is exactly the quantum break time. This is because a pure $\rho^{(1)}$ represents a condensate-like state with all bosons in one orbital. This state can be completely described by a classical field representing the wave function of the relevant orbital. Therefore as soon as $\rho^{(1)}$ develops a large entropy, the gas can no longer be expected to have a classical description. 

The one-particle density matrix may be diagonalized
\begin{equation}
\rho^{(1)} = \sum_i \lambda_i \ket{\Phi_i}\bra{\Phi_i} \, .
\end{equation}
with eigenvectors $\ket{\Phi_i}$, $\lambda_i$ and eigenvalues $\rho^{(1)}(\Psi)$. 

A true BE condensate state $\ket{\Psi_{BE}}$ is characterized by possessing one eigenvalue $\lambda_{max}=O(1)$ with the sum of all other eigenvalues suppressed as $1/N$. If a many-body ground state is of this type, we will say that the system is a BE condensate. 

In the limit $N  \to  \infty$ the corresponding reduced one-particle density matrix $\rho^{(1)}$ defines a pure state $\ket{\Phi_{GP}}$ in the one-particle Hilbert space, which is the eigenvector corresponding to the unique maximal eigenvalue. 
The BE many-body state corresponds to having all the $N$ constituents in the same state $\ket{\Phi_{GP}}$. The wave function $\Phi_{GP}(x,t)$ of this one-particle state is the Gross-Pitaevskii wave function and its evolution is described by the Gross-Pitaevskii equation (\ref{eq:gross-pitaevskii}).

For finite $N$ and finite $gN$, the Gross-Pitaevskii equation is never exact. In fact, any exact BE condensate state will, by quantum mechanical time evolution, deplete. This is reflected by the fact that the other eigenvalues of $\rho^{(1)}$ grow. In what follows, we are interested in tracking precisely this growth for some concrete initial conditions, as this allows us to quantify how quickly the Gross-Pitaevskii description breaks down. 

Under these conditions the quantum break time $t_b$ appears as the time in which the difference between the exact many-body evolution and the mean field time evolution surpasses a threshold value. Note that the scaling of $t_b$ with $N$ is independent of the choice of threshold value, therefore rendering it effectively arbitrary for our purposes.

Before going into more concrete details let us briefly discuss the physical meaning of this timescale. Let us denote by $\rho^{(1)}(t)$ the exact many-body evolution of the reduced density matrix, whereas by $\rho^{(1)}_{GP}(t)$ we label the mean field GP time evolution for the same initial conditions at $t=0$. Since $\rho^{(1)}_{GP}(t)$ is a pure state, we can use as a measure of the difference with respect to $\rho^{(1)}(t)$ the entanglement entropy $S(\rho^{(1)}(t))$. We will define $t_b$ as the time needed to reach a certain threshold entropy. This time will generically depend both on the initial condition as well as  on the number $N$ of constituents. 

The potential growth of the entanglement with time means that the one-particle density matrix is losing {\it quantum coherence}. On the other hand, and from the point of view of the many body wave function, this loss of quantum coherence is reflected in the form of {\it quantum depletion}, i.e. in the growth of the number of constituents that are not in the condensate state. Note,  that since at the time $t_b$ the number of constituents away from the condensate is significant, this time also sets the limit of applicability of the Bogolyubov approximation.

For regular quantum systems we can expect the time $t_b$ to depend on $N$ as some power \cite{Ehrenfest}.
However, as we will show, some attractive BE condensates exhibit a quantum breaking time scaling with $N$ as $t_b \,  \sim \, \log N$ 
i.e., they generate entanglement in a time depending on the effective Planck constant as $\log(1/\hbar)$.

In this sense BE condensates -- under those conditions -- effectively behave as {\it fast scramblers}. Hence our task will be, on one side to identify the above conditions and on the other side to relate those fast scrambler BE condensates with the sort of BE condensates we have put forward as microscopic portraits of black holes.

\section{Scrambling and Quantumness in BE Condensates}
A necessary condition for having a quantum break time $t_b$ scaling like $\log N$ for some initial many body state $\Psi_0$ is the exponential growth with time of small fluctuations $\delta \Psi(t)$ where $\Psi = \Psi_0 + \delta \Psi$. In linear approximation the equation controlling $\delta \Psi$ is the Bogolyubov-De-Gennes equation. As discussed above, a significant departure from the mean field approximation as well as generation of entanglement for the reduced one particle density matrix requires a growth in time of the depleted i.e of the non-condensed particles. Nicely enough the equations controlling the growth of depleted particles are the same as the ones controlling the small fluctuations of the Gross-Pitaevskii equation and therefore we can translate the problem of finding a time $t_b$ scaling like $\log N$ into the simpler problem of the \emph{stability of the Gross-Pitaevskii equation}. For a detailed discussion and the related technicalities, see \cite{castin1998low}.

We can understand the short break time more concretely if we think about the difference between the exact evolution and the mean field evolution as the addition of a small perturbation to the exact Hamiltonian.  Since an unstable system is exponentially sensitive to perturbations of the Hamiltonian then the time for the evolution of states to differ substantially is very short.
The instability is controlled by the Lyapunov exponent $\lambda$, while the preexponential factor will depend on the size of the perturbation. The quantum break time is the time when this becomes important, so we can naturally expect it to scale like
$t_b \sim \lambda^{-1} \log N$.

\section{Numerical Analysis\label{numerics}}

\subsection{Quantum Break Time of One Dimensional Condensates}

In this section we will verify the logarithmic quantum break time numerically for the (1+1)-d Bose condensate. 

The theory (\ref{eq:hamiltonian}) in $1+1$ dimensions undergoes a quantum phase transition for $gN = 1$. When surpassing the critical coupling, the homogeneous state becomes dynamically unstable. 

As we expect the black hole to lie at such a point of instability, due to its collapse going in hand with Hawking evaporation, we will model the behavior of the black hole by considering the homogeneous state past the point of quantum phase transition.

We consider $gN > 1$ and prepare as initial condition a perfect condensate in the homogeneous one-particle orbital . The linear stability analysis (simply expanding the classical Hamiltonian (\ref{eq:hamiltonian}) around a the background) at once indicates an instability: the energy of the first Bogolyubov mode becomes imaginary; its magnitude corresponds to $\lambda$, the Lyapunov coefficient for the unstable direction. 

Note that this setup may be interpreted as preparing the system in a supercooled phase. Or as the result of a quench across the phase transition, suddenly increasing the coupling from $gN = 0$ to $gN > 1$. The system finds itself in a classically instable configuration and quantum fluctuations ensure that a rapid depletion of the condensate and simultaneous entanglement generation take place. 

Would we evolve the same initial state for $gN < 1$, very little entanglement would be generated (because it overlaps with very few energy eigenstates there) and the relevant timescale of evolution would not scale logarithmically in $N$ (as can be checked by studying the spectrum).

Decomposition of $\phi$ in terms of annihilation and creation operators
\begin{equation}
	\label{psidec} 
	\hat\phi = \frac{1}{\sqrt{L_b}}\sum_{k=-\infty}^{\infty} \hat{a}_k e^{i k x} \,,
\end{equation}
leads to the more convenient form for (\ref{eq:hamiltonian})\footnote{For improved readability, we have now set $\hbar = 2m = V = 1$}
\begin{equation}
\label{eq:momhamilt}
\hat{H} = \sum_{k=-\infty}^\infty k^2 \hat{a}_k^\dagger \hat{a}_k - \frac{g}{4}\sum_{k,l,m = -\infty}^\infty \hat{a}_k^\dagger \hat{a}_l^\dagger \hat{a}_{m+k} \hat{a}_{l-m}
\end{equation}
Bogolyubov diagonalization around the homogeneous background $\phi_\text{hom} = \sqrt{N}$ yields for the energy of the first Bogolyubov mode \cite{Kanamoto:2002xy, Dvali:2012wq, Flassig:2012re}
\begin{equation}
\label{eq:bogfreq}
\epsilon_1 = \sqrt{1-gN}
\end{equation}
Parametrizing the effective coupling as $gN = 1 + \delta$, we obtain $\epsilon_1 = i \sqrt{\delta}$.
Applying the above argument, we therefore expect the system to break from mean field on a timescale $t_\mathrm{break} \sim \Im(\epsilon_1)^{-1} \log N \sim \sqrt{\delta}^{-1} \log N$. The argument of the logarithm is proportional to $N$ because 
 the action of the mean field solution scales as $S \sim N$ for fixed $gN$.

Within this setup, the departure from classical evolution is expected to go in hand with the generation of large entanglement. This allows us to identify the quantum break time directly with the scrambling time.

Since we are interested in finite N effects in a regime where we expect semi-classical methods to fail, we will use a method not relying on any kind of perturbation theory. We will diagonalize the Hamiltonian (\ref{eq:momhamilt}) explicitly. Then, in order to time evolve the homogeneous Hartree state 
\begin{equation}
|\phi_\text{hom}\rangle = (\hat{a}_0^\dagger)^N|0\rangle ,
\end{equation}
 we will project $|\phi_\text{hom}\rangle$ onto energy eigenstates and apply the time evolution operator  $U(t) = \exp{(i H t)}$ on the state. Finally, we project the time evolved state onto a k-particle subspace and compute the von Neumann entropy 
\begin{eqnarray}
S_{1} &=& -\tr \rho_{1} \log \rho_{1} \nonumber \\
(\rho_{1})_{ij} &=& \bra{\phi_\text{hom}}\hat{a}_i^\dagger\hat{a}_j\ket{\phi_\text{hom}}
\end{eqnarray}
as a function of time. 

In order to make this task computationally feasible we will make use of several properties of the system \cite{Kanamoto:2002xy}. Since the Hamiltonian is translationally invariant and number conserving we can restrict ourselves to fixed total momentum and fixed total particle number. In our case, only the total momentum zero sector is relevant, since this contains the homogeneous state. Furthermore, from the Bogolyubov analysis we see that the modes with $k>2$ have a fairly large gap for $gN$ not much bigger than $1$. Therefore, we can truncate the momentum modes $l$ we take into account to $l={-1,0,1}$.

\begin{figure}[t]
  \centering
  \includegraphics[width=\linewidth]{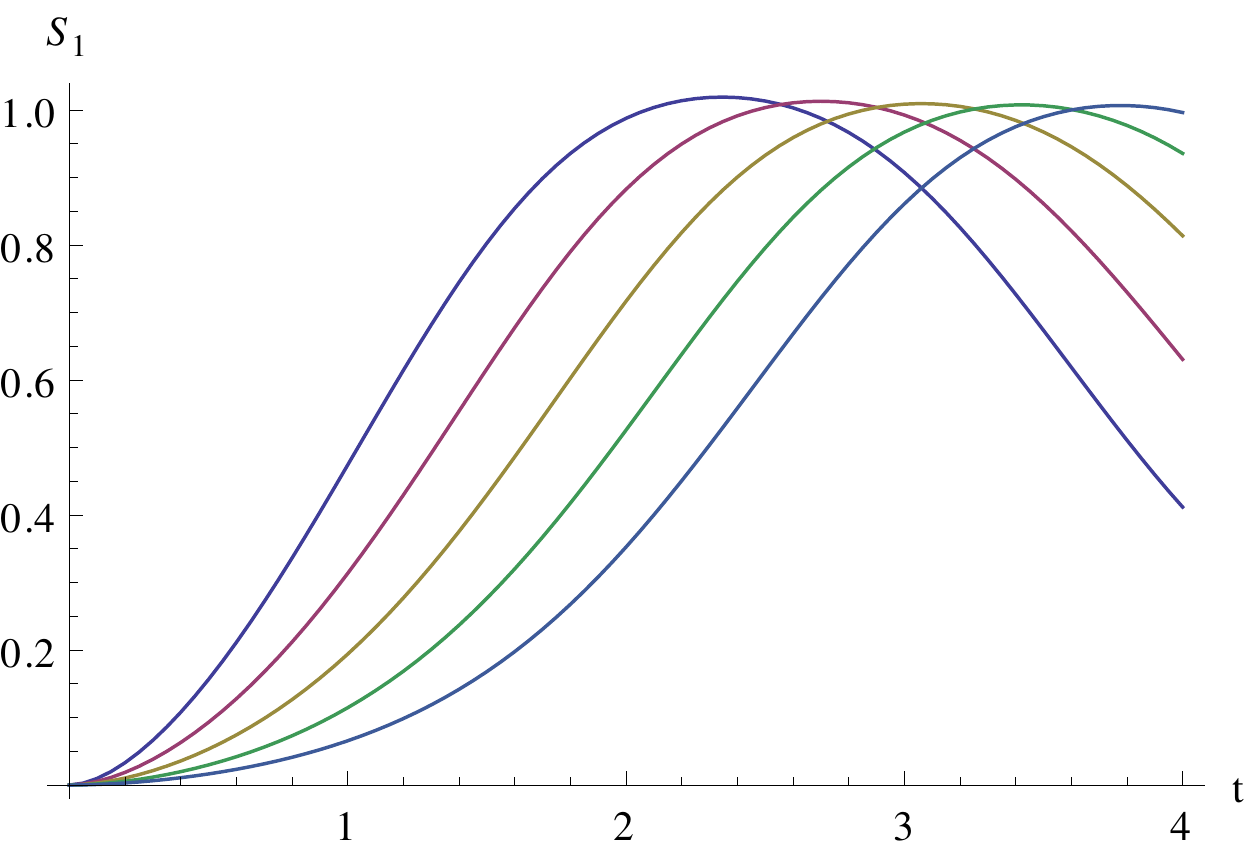}
  \caption{One particle entanglement entropy as a function of time for $N = 16,32,64,128$ and $256$.}
    \label{fig:enttime}
\end{figure}

\begin{figure}[t]
  \centering
\includegraphics[scale=1]{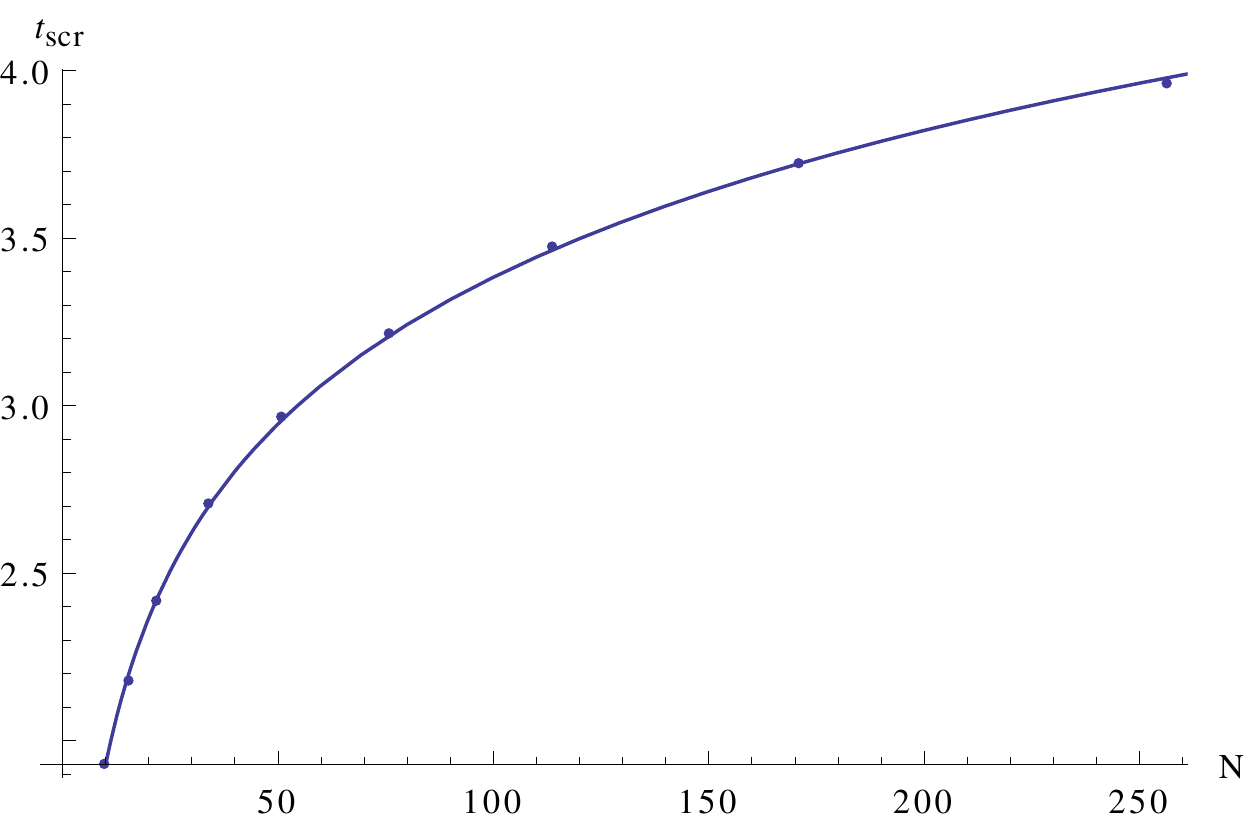} 
  \caption{Quantum break time as a function of $N$.}
    \label{fig:breaktN}
\end{figure}

In Fig. \ref{fig:enttime} we plot $S_1$ as functions of time for different values of N. In order to see the break time, we evaluate the time when $S_1$ is higher than some threshold value $S_{th}$. We plot this time as a function of particle number N in Fig. \ref{fig:breaktN}, where the solid line is the result of fitting a logarithm to the data points. This clearly shows a logarithmic break time. 

\begin{figure}[t]
  \centering
\includegraphics[scale=1]{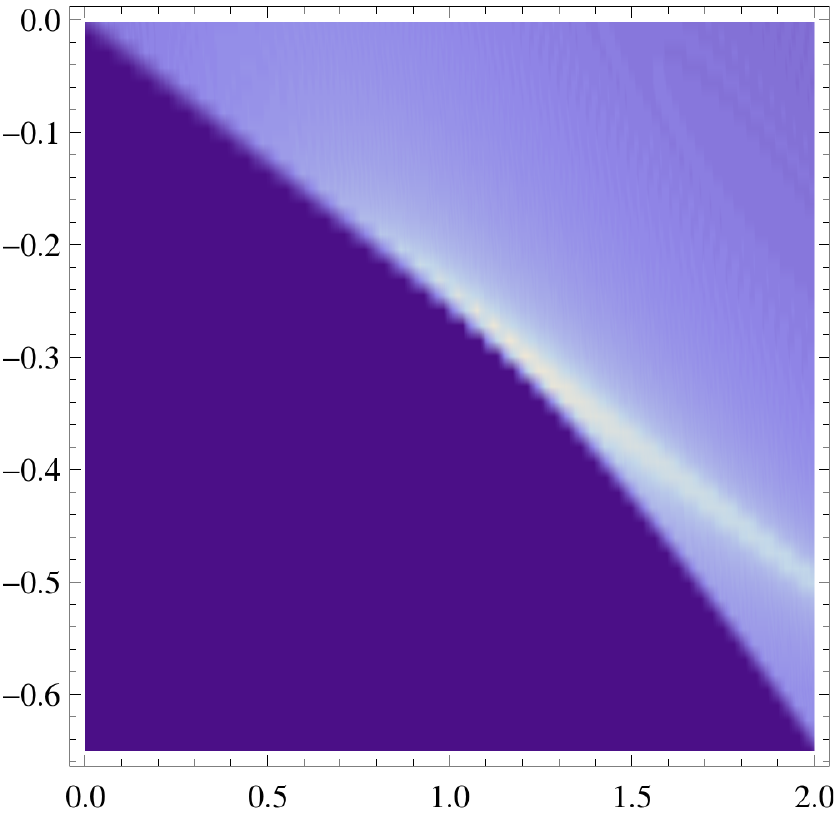} 
  \caption{Density of states as a function of $gN$ and $E/N$ for N=1500.}
    \label{fig:densstates}
\end{figure}

A clearer understanding for the observed behavior emerges if we look at the density of states. In Fig. \ref{fig:densstates} we show a plot of the density of states in the zero-momentum sector for given energy and coupling. It can be clearly observed that there is a large density of states for low energies near the phase transition, which is due to the light Bogolyubov mode appearing at the quantum critical point. Furthermore, we clearly see a band of a high density of states for large couplings. The state we time-evolve in the numerics overlaps only with the modes in this band. We have checked that the density of states in this band varies logarithmically with N, i.e. the gap $\Delta$ between states in this band will typically go as
\begin{equation}
	\Delta\sim 1/(\lambda \log N).
\end{equation}
Given that the time scale for the time evolution will be set by this gap we naturally see the logarithmic break time emerging.

\subsection{Three Dimensional Condensates and Connection with Black Hole}

\begin{figure}[t]
  \centering
  \includegraphics[width=\linewidth,angle=0.]{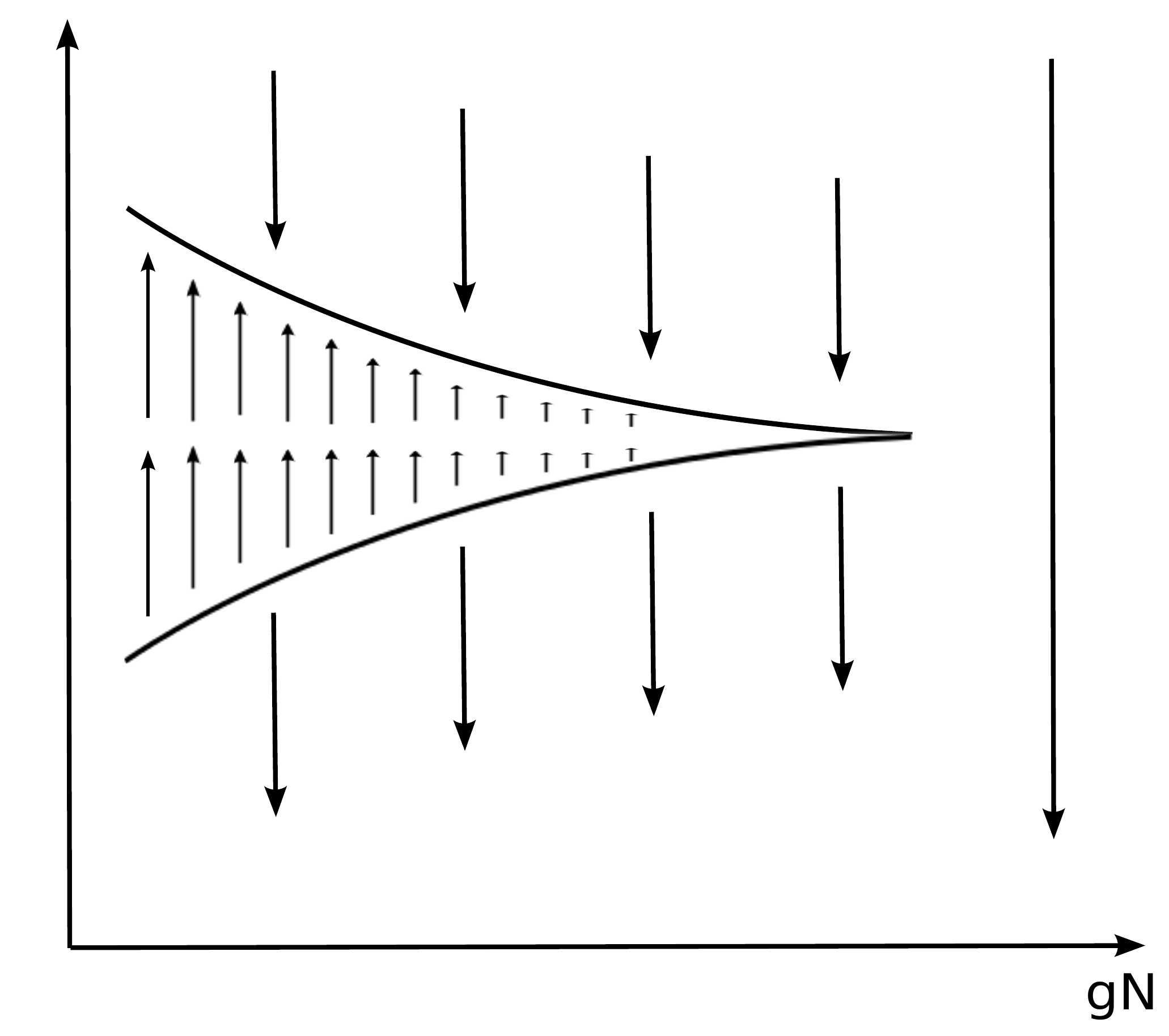}
  \caption{Phase diagram for the three-dimensional condensate. For small $gN$ two solutions exist; one is stable while the other one is unstable. At the critical point, both solutions disappear.}
    \label{fig:3dPhaseDiagram}
\end{figure}

In the previous section, we have studied a Bose condensate in one spatial dimension as a prototype model. In that case it was viable to perform numerical simulations of the quantum time evolution. For an attractive Bose condensate, one dimension is special however insofar as the classical GP system has a well defined lowest energy configuration after the phase transition - the bright soliton. In higher dimensions, however, there is no classical solution in the would-be solitonic phase. Instead when increasing the effective coupling $gN$ past $1$, the stable lowest energy solution of the Gross-Pitaevskii equation and another (unstable) solution disappear together in a saddle node bifurcation \cite{PhysRevA.56.1424} (see a sketch of the phase diagram in Fig. \ref{fig:3dPhaseDiagram}\footnote{This can also be understood intuitively from Fig.\ref{fig:escal} and Eq.(\ref{eq:escal}). The two solutions for small $gN$ correspond to the maximum of the energy functional and the infinitely stretched condensate. For large $gN$, no stable points exist. This analysis assumes the presence of a trapping potential. As we will argue below, this is in close analogy to the black hole.}). Thus, while we willfully prepared an unstable initial state for the (1+1)-d bosons, when a perfectly stable ground state was available, in (3+1)-d it is inevitable to enter the instability when going past the bifurcation point. 

It is precisely this instability that we believe to be responsible for the fast scrambling of information in black holes.

There, the relevant coupling controlling the mean field approximation is $gN$ with $g=\frac{L_P^2}{l^2}$ for $l$ the wave length of the constituent gravitons. In the weak coupling regime $gN<1$ the condensate cannot be self-sustained and we should therefore imagine some external trapping potential that sets the wavelength of the constituent gravitons. The many body wave function is a stretched condensate in the corresponding trap. At the critical point $gN=1$ the system of gravitons becomes self-sustained in the sense that the quantum pressure compensates the gravitational attraction. However, although at this point we can satisfy the virial condition of self-sustainability, the system is not stable in the mean field approximation and will tend to collapse - reducing its size and consequently decreasing the typical wavelength of the constituent gravitons. As we have elaborated, this mean field picture dramatically changes once we take appropriately into account $1/N$ quantum effects. 
Based on our prototypes, we expect the quantum evolution to break from mean field in a time $O(R \log N)$. This is reflected in the generation of large entanglement entropy for the corresponding one particle density matrix as a function of $gN$.

The evolution of black holes is different from that of laboratory condensates because of Hawking evaporation. While collapse usually puts a condensate off the critical point, this is prevented by the decrease of the number of gravitons $N$. 
As the condition of instability persists along the collapse, we also expect larger-$k$-density matrices to be efficiently scrambled.

\section{Summary and Outlook}

    The purpose of this note was to stress that the properties of unstable Bose-Einstein condensates 
  are crucial in understanding the efficient generation of quantum entanglement and scrambling.

The very conservative assumption of our work lies in modeling black holes as many-body quantum systems governed by weakly-coupled IR gravity. The semi-classical one-particle collective behavior
appears as a consequence of the many-body system being in a BE condensate state. Quantum fluctuations relative to this state are measured by $1/N$ with $N$ being the number of graviton constituents (and, equivalently, the BH action in Planck units). Some special features of BHs, as, for instance, fast scrambling, are understood in this frame as the reflection of a logarithmic quantum break time. 

These observations provide the clue for solving some recalcitrant BH paradoxes. In particular, the assumption of purity of the final evaporation state seems to lead to strong departures from semi-classicality at least in Page's time \cite{Page:1993wv}, meaning that a breakdown of semi-classicality takes place after this time  irrespective of the size of the black hole.  This is very puzzling, since naively one expects the
semi-classical approximation to be valid for large macroscopic black holes. 
The approach to these sort of puzzles that we can extract from the present work lies in identifying the root of this breakdown of semi-classicality in the existence of a logarithmic quantum break time. 
Because BHs are unstable BECs,  the quantum evolution takes over much sooner then what would be naively expected. 

As a marginal comment let us just note that the quantum time coordinate $\lambda t$, with $\lambda$ the Lyapunov exponent, is the natural candidate for the Rindler time. This leads to a potential connection between the physics of the time-coordinate  inside the black hole and the entanglement flow for the reduced density matrix.  

Finally,  it would be very interesting  to study some of the phenomena  discussed in this work, in particular the appearance of logarithmic quantum break time, for realistic Bose-Einstein  condensates in the laboratory. This would give an exciting prospect of simulating some aspects of quantum black hole physics in the labs.

\section*{Acknowledgments}
We thank Sarah Folkerts ,Andre Franca,Valentino Foit and Andrei Khmelnitsky for comments and discussions.
The work of G.D. was supported in part by Humboldt Foundation under Alexander von Humboldt Professorship,  by European Commission  under 
the ERC advanced grant 226371,   by TRR 33 \textquotedblleft The Dark
Universe\textquotedblright\   and  by the NSF grant PHY-0758032. 
The work of C.G. was supported in part by Humboldt Foundation and by Grants: FPA 2009-07908, CPAN (CSD2007-00042) and HEPHACOS P-ESP00346.
D.F., A.P. and N.W. were supported by the Humboldt Foundation.

 \bibliography{biblio}{}

\providecommand{\href}[2]{#2}\begingroup\raggedright\begin{thebibliography}{10}

\bibitem{Dvali:2011aa}
G.~Dvali and C.~Gomez, ``{Black Hole's Quantum N-Portrait},''
  \href{http://dx.doi.org/10.1002/prop.201300001}{{\em Fortsch.Phys.}
  {\bfseries 61} (2011) 742--767},
\href{http://arxiv.org/abs/1112.3359}{{\ttfamily arXiv:1112.3359 [hep-th]}}.
%%CITATION = ARXIV:1112.3359;%%.

\bibitem{Dvali:2012en}
G.~Dvali and C.~Gomez, ``{Black Holes as Critical Point of Quantum Phase
  Transition},''
\href{http://arxiv.org/abs/1207.4059}{{\ttfamily arXiv:1207.4059 [hep-th]}}.
%%CITATION = ARXIV:1207.4059;%%.

\bibitem{Dvali:2012wq}
G.~Dvali and C.~Gomez, ``{Black Hole Macro-Quantumness},''
\href{http://arxiv.org/abs/1212.0765}{{\ttfamily arXiv:1212.0765 [hep-th]}}.
%%CITATION = ARXIV:1212.0765;%%.

\bibitem{Dvali:2012rt}
G.~Dvali and C.~Gomez, ``{Black Hole's 1/N Hair},''
  \href{http://dx.doi.org/10.1016/j.physletb.2013.01.020}{{\em Phys.Lett.}
  {\bfseries B719} (2013) 419--423},
\href{http://arxiv.org/abs/1203.6575}{{\ttfamily arXiv:1203.6575 [hep-th]}}.
%%CITATION = ARXIV:1203.6575;%%.

\bibitem{Flassig:2012re}
D.~Flassig, A.~Pritzel, and N.~Wintergerst, ``{Black Holes and Quantumness on
  Macroscopic Scales},''
  \href{http://dx.doi.org/10.1103/PhysRevD.87.084007}{{\em Phys.Rev.}
  {\bfseries D87} (2013) 084007},
\href{http://arxiv.org/abs/1212.3344}{{\ttfamily arXiv:1212.3344 [hep-th]}}.
%%CITATION = ARXIV:1212.3344;%%.

\bibitem{berkhahn2013microscopic}
F.~Berkhahn, S.~M\"uller, F.~Niedermann, and R.~Schneider, ``{Microscopic
  Picture of Non-Relativistic Classicalons},''
\href{http://arxiv.org/abs/1302.6581}{{\ttfamily arXiv:1302.6581 [hep-th]}}.
%%CITATION = ARXIV:1302.6581;%%.

\bibitem{Mueck:2013mha}
W.~Mueck, ``{On the number of soft quanta in classical field configurations},''
\href{http://arxiv.org/abs/1306.6245}{{\ttfamily arXiv:1306.6245 [hep-th]}}.
%%CITATION = ARXIV:1306.6245;%%.

\bibitem{Binetruy:2012kx}
P.~Binetruy, ``{Vacuum energy, holography and a quantum portrait of the visible
  Universe},''
\href{http://arxiv.org/abs/1208.4645}{{\ttfamily arXiv:1208.4645 [gr-qc]}}.
%%CITATION = ARXIV:1208.4645;%%.

\bibitem{Casadio:2013hja}
R.~Casadio and A.~Orlandi, ``{Quantum Harmonic Black Holes},''
\href{http://arxiv.org/abs/1302.7138}{{\ttfamily arXiv:1302.7138 [hep-th]}}.
%%CITATION = ARXIV:1302.7138;%%.

\bibitem{Dvali:2012uq}
G.~Dvali, C.~Gomez, and D.~Lust, ``{Black Hole Quantum Mechanics in the
  Presence of Species},'' \href{http://dx.doi.org/10.1002/prop.201300002}{{\em
  Fortsch.Phys.} {\bfseries 61} (2013) 768--778},
\href{http://arxiv.org/abs/1206.2365}{{\ttfamily arXiv:1206.2365 [hep-th]}}.
%%CITATION = ARXIV:1206.2365;%%.

\bibitem{strominger1996microscopic}
A.~Strominger and C.~Vafa, ``{Microscopic origin of the Bekenstein-Hawking
  entropy},'' \href{http://dx.doi.org/10.1016/0370-2693(96)00345-0}{{\em
  Phys.Lett.} {\bfseries B379} (1996) 99--104},
\href{http://arxiv.org/abs/hep-th/9601029}{{\ttfamily arXiv:hep-th/9601029
  [hep-th]}}.
%%CITATION = HEP-TH/9601029;%%.

\bibitem{Banks:1996vh}
T.~Banks, W.~Fischler, S.~Shenker, and L.~Susskind, ``{M theory as a matrix
  model: A Conjecture},''
  \href{http://dx.doi.org/10.1103/PhysRevD.55.5112}{{\em Phys.Rev.} {\bfseries
  D55} (1997) 5112--5128},
\href{http://arxiv.org/abs/hep-th/9610043}{{\ttfamily arXiv:hep-th/9610043
  [hep-th]}}.
%%CITATION = HEP-TH/9610043;%%.

\bibitem{Banks:1997hz}
T.~Banks, W.~Fischler, I.~R. Klebanov, and L.~Susskind, ``{Schwarzschild black
  holes from matrix theory},''
  \href{http://dx.doi.org/10.1103/PhysRevLett.80.226}{{\em Phys.Rev.Lett.}
  {\bfseries 80} (1998) 226--229},
\href{http://arxiv.org/abs/hep-th/9709091}{{\ttfamily arXiv:hep-th/9709091
  [hep-th]}}.
%%CITATION = HEP-TH/9709091;%%.

\bibitem{Banks:1997tn}
T.~Banks, W.~Fischler, I.~R. Klebanov, and L.~Susskind, ``{Schwarzschild black
  holes in matrix theory. 2.},'' {\em JHEP} {\bfseries 9801} (1998) 008,
\href{http://arxiv.org/abs/hep-th/9711005}{{\ttfamily arXiv:hep-th/9711005
  [hep-th]}}.
%%CITATION = HEP-TH/9711005;%%.

\bibitem{mathur2005fuzzball}
S.~D. Mathur, ``{The Fuzzball proposal for black holes: An Elementary
  review},'' \href{http://dx.doi.org/10.1002/prop.200410203}{{\em
  Fortsch.Phys.} {\bfseries 53} (2005) 793--827},
\href{http://arxiv.org/abs/hep-th/0502050}{{\ttfamily arXiv:hep-th/0502050
  [hep-th]}}.
%%CITATION = HEP-TH/0502050;%%.

\bibitem{Hayden:2007cs}
P.~Hayden and J.~Preskill, ``{Black holes as mirrors: Quantum information in
  random subsystems},''
  \href{http://dx.doi.org/10.1088/1126-6708/2007/09/120}{{\em JHEP} {\bfseries
  0709} (2007) 120},
\href{http://arxiv.org/abs/0708.4025}{{\ttfamily arXiv:0708.4025 [hep-th]}}.
%%CITATION = ARXIV:0708.4025;%%.

\bibitem{Sekino:2008he}
Y.~Sekino and L.~Susskind, ``{Fast Scramblers},''
  \href{http://dx.doi.org/10.1088/1126-6708/2008/10/065}{{\em JHEP} {\bfseries
  0810} (2008) 065},
\href{http://arxiv.org/abs/0808.2096}{{\ttfamily arXiv:0808.2096 [hep-th]}}.
%%CITATION = ARXIV:0808.2096;%%.

\bibitem{Maldacena:2001kr}
J.~M. Maldacena, ``{Eternal black holes in anti-de Sitter},'' {\em JHEP}
  {\bfseries 0304} (2003) 021,
\href{http://arxiv.org/abs/hep-th/0106112}{{\ttfamily arXiv:hep-th/0106112
  [hep-th]}}.
%%CITATION = HEP-TH/0106112;%%.

\bibitem{Hawking:2005kf}
S.~Hawking, ``{Information loss in black holes},''
  \href{http://dx.doi.org/10.1103/PhysRevD.72.084013}{{\em Phys.Rev.}
  {\bfseries D72} (2005) 084013},
\href{http://arxiv.org/abs/hep-th/0507171}{{\ttfamily arXiv:hep-th/0507171
  [hep-th]}}.
%%CITATION = HEP-TH/0507171;%%.

\bibitem{Veneziano:2012yj}
G.~Veneziano, ``{Quantum hair and the string-black hole correspondence},''
  \href{http://dx.doi.org/10.1088/0264-9381/30/9/092001}{{\em
  Class.Quant.Grav.} {\bfseries 30} (2013) 092001},
\href{http://arxiv.org/abs/1212.2606}{{\ttfamily arXiv:1212.2606 [hep-th]}}.
%%CITATION = ARXIV:1212.2606;%%.

\bibitem{Brustein:2013qma}
R.~Brustein and A.~Medved, ``{Restoring predictability in semiclassical
  gravitational collapse},''
\href{http://arxiv.org/abs/1305.3139}{{\ttfamily arXiv:1305.3139 [hep-th]}}.
%%CITATION = ARXIV:1305.3139;%%.

\bibitem{Brustein:2013xga}
R.~Brustein and A.~Medved, ``{Semiclassical black holes expose forbidden
  charges and censor divergent densities},''
\href{http://arxiv.org/abs/1302.6086}{{\ttfamily arXiv:1302.6086 [hep-th]}}.
%%CITATION = ARXIV:1302.6086;%%.

\bibitem{Casadio:2013aua}
R.~Casadio and F.~Scardigli, ``{Horizon wave-function for single localized
  particles: GUP and quantum black hole decay},''
\href{http://arxiv.org/abs/1306.5298}{{\ttfamily arXiv:1306.5298 [gr-qc]}}.
%%CITATION = ARXIV:1306.5298;%%.

\bibitem{Susskind2}
L.~Susskind, ``{Addendum to Fast Scramblers},''
\href{http://arxiv.org/abs/1101.6048}{{\ttfamily arXiv:1101.6048 [hep-th]}}.
%%CITATION = ARXIV:1101.6048;%%.

\bibitem{Lashkari:2011yi}
N.~Lashkari, D.~Stanford, M.~Hastings, T.~Osborne, and P.~Hayden, ``{Towards
  the Fast Scrambling Conjecture},''
  \href{http://dx.doi.org/10.1007/JHEP04(2013)022}{{\em JHEP} {\bfseries 1304}
  (2013) 022},
\href{http://arxiv.org/abs/1111.6580}{{\ttfamily arXiv:1111.6580 [hep-th]}}.
%%CITATION = ARXIV:1111.6580;%%.

\bibitem{Asplund:2011qj}
C.~Asplund, D.~Berenstein, and D.~Trancanelli, ``{Evidence for fast
  thermalization in the plane-wave matrix model},''
  \href{http://dx.doi.org/10.1103/PhysRevLett.107.171602}{{\em Phys.Rev.Lett.}
  {\bfseries 107} (2011) 171602},
\href{http://arxiv.org/abs/1104.5469}{{\ttfamily arXiv:1104.5469 [hep-th]}}.
%%CITATION = ARXIV:1104.5469;%%.

\bibitem{Barbon:2011pn}
J.~L. Barbon and J.~M. Magan, ``{Chaotic Fast Scrambling At Black Holes},''
  \href{http://dx.doi.org/10.1103/PhysRevD.84.106012}{{\em Phys.Rev.}
  {\bfseries D84} (2011) 106012},
\href{http://arxiv.org/abs/1105.2581}{{\ttfamily arXiv:1105.2581 [hep-th]}}.
%%CITATION = ARXIV:1105.2581;%%.

\bibitem{altland2012quantum}
A.~{Altland} and F.~{Haake}, ``{Quantum Chaos and Effective Thermalization},''
  \href{http://dx.doi.org/10.1103/PhysRevLett.108.073601}{{\em Physical Review
  Letters} {\bfseries 108} no.~7, (Feb., 2012) 073601},
  \href{http://arxiv.org/abs/1110.1270}{{\ttfamily arXiv:1110.1270 [nlin.CD]}}.

\bibitem{Kanamoto:2002xy}
R.~Kanamoto, H.~Saito, and M.~Ueda, ``{Quantum phase transition in
  one-dimensional Bose-Einstein condensates with attractive interactions},''
  \href{http://dx.doi.org/10.1103/PhysRevA.67.013608}{{\em Phys. Rev. A}
  {\bfseries 67} (Jan, 2003) 013608}.

\bibitem{Ehrenfest}
P.~Ehrenfest, ``{Bemerkung \"uber die angen\"aherte G\"ultigkeit der
  klassischen Mechanik innerhalb der Quantenmechanik},''
  \href{http://dx.doi.org/10.1007/BF01329203}{{\em Zeitschrift f\"ur Physik}
  {\bfseries 45} no.~7-8, (1927) 455--457}.

\bibitem{castin1998low}
Y.~{Castin} and R.~{Dum}, ``{Low-temperature Bose-Einstein condensates in
  time-dependent traps: Beyond the U(1) symmetry-breaking approach},''
  \href{http://dx.doi.org/10.1103/PhysRevA.57.3008}{{\em Phys. Rev. A}
  {\bfseries 57} (Apr, 1998) 3008--3021}.

\bibitem{PhysRevA.56.1424}
V.~M. P\'erez-Gar\'cia, H.~Michinel, J.~I. Cirac, M.~Lewenstein, and P.~Zoller,
  ``{Dynamics of Bose-Einstein condensates: Variational solutions of the
  Gross-Pitaevskii equations},''
  \href{http://dx.doi.org/10.1103/PhysRevA.56.1424}{{\em Phys. Rev. A}
  {\bfseries 56} (Aug, 1997) 1424--1432}.

\bibitem{Page:1993wv}
D.~N. Page, ``{Information in black hole radiation},''
  \href{http://dx.doi.org/10.1103/PhysRevLett.71.3743}{{\em Phys.Rev.Lett.}
  {\bfseries 71} (1993) 3743--3746},
\href{http://arxiv.org/abs/hep-th/9306083}{{\ttfamily arXiv:hep-th/9306083
  [hep-th]}}.
%%CITATION = HEP-TH/9306083;%%.

\bibitem{zurek1994decoherence}
W.~H. {Zurek} and J.~P. {Paz}, ``{Decoherence, chaos, and the second law},''
  \href{http://dx.doi.org/10.1103/PhysRevLett.72.2508}{{\em Physical Review
  Letters} {\bfseries 72} (Apr., 1994) 2508--2511},
  \href{http://arxiv.org/abs/arXiv:gr-qc/9402006}{{\ttfamily
  arXiv:gr-qc/9402006}}.

\bibitem{Anglin:2001yz}
J.~R. Anglin and A.~Vardi, ``{Dynamics of a two-mode Bose-Einstein condensate
  beyond mean-field theory},''
  \href{http://dx.doi.org/10.1103/PhysRevA.64.013605}{{\em Phys. Rev. A}
  {\bfseries 64} (May, 2001) 013605}.

\end{thebibliography}\endgroup
\bibliographystyle{utphys}

\begin{appendices}
\section{Instability and Quantum Break Time \label{husimi-argument}}
The question of how long a mean field (i.e.\ classical) trajectory faithfully reproduces the quantum evolution of a dynamical system has been studied a very long time ago \cite{Ehrenfest}. Only much later however has it been noticed, that under certain circumstances, the quantum evolution can deviate from mean field in sub-polynomial time. Good arguments have been given \cite{zurek1994decoherence} that where the classical phase space of a system exhibits a dynamical instability, i.e. a Lyapunov exponent $\lambda > 1$, the quantum dynamics will deviate from mean field after a time that goes like
\begin{equation}
	t_\mathrm{break} = \lambda^{-1} \log(S / \hbar)
	\label{eq:quantum-break-time}
\end{equation} 
where $S$ is the typical action.

\subsection{Local Instability Argument}
The general argument (following \cite{zurek1994decoherence}) that leads to the logarithmic break time can be summarized as follows: Assume that the classical phase space of the system contains a region with a local Lyapunov exponent $\lambda > 0$. For the sake of the argument, let us represent every pure quantum state $| \psi \rangle$ as a Glauber $Q$ or Husimi quasi-probability distribution\footnote{It is similar to the better known Wigner quasi-probability function, but has some properties that make it favorable for the study of chaotic systems.} on phase space
\begin{equation}
	Q_\psi(\alpha) = \frac{1}{\pi} \left| \langle \alpha | \psi \rangle \right|^2 \,,
\end{equation} 
where the $| \alpha \rangle$ form an overcomplete basis of coherent states. As one would expect for a real probability distribution, the $Q$ distribution moves along with the classical trajectories. There are however intrinsically quantum terms that contain additional derivatives and act diffusively on phase space. 

Imagine that we initially prepare a close analog to classical state - a coherent state - and localize it in the unstable region of phase space. The $Q$ distribution, initially well localized, will be stretched in the unstable direction. However, because Hamiltonian flows are volume preserving, there must also be a ``stable'' direction with a local Lyapunov exponent $- \lambda$. The $Q$ distribution is exponentially compressed in the stable direction. When its width gets smaller than a given phase space distance (that involves $\hbar$), the diffusive quantum terms become important. From that point on, the quantum time evolution departs even from the physics of a classical phase space ensemble. The time scale for the departure naturally goes like 
\begin{equation}
	t_\mathrm{break} = \lambda^{-1} \log(1/\hbar).
\end{equation}
It can be argued that the dimensionless ratio in the exponent should be $S/\hbar$ with $S$ being the typical action \cite{Ehrenfest}. The quantum break time $t_\mathrm{break}$ is also referred to as Ehrenfest time in the literature on quantum chaos.

The quick break time has been explicitly verified numerically, e.g.\ in tractable two level systems that are well motivated experimentally \cite{Anglin:2001yz}.

\subsection{Quantum Break Time for a Wave Packet}
\begin{figure}[p]
  \centering
  \includegraphics[width=\linewidth,angle=0.]{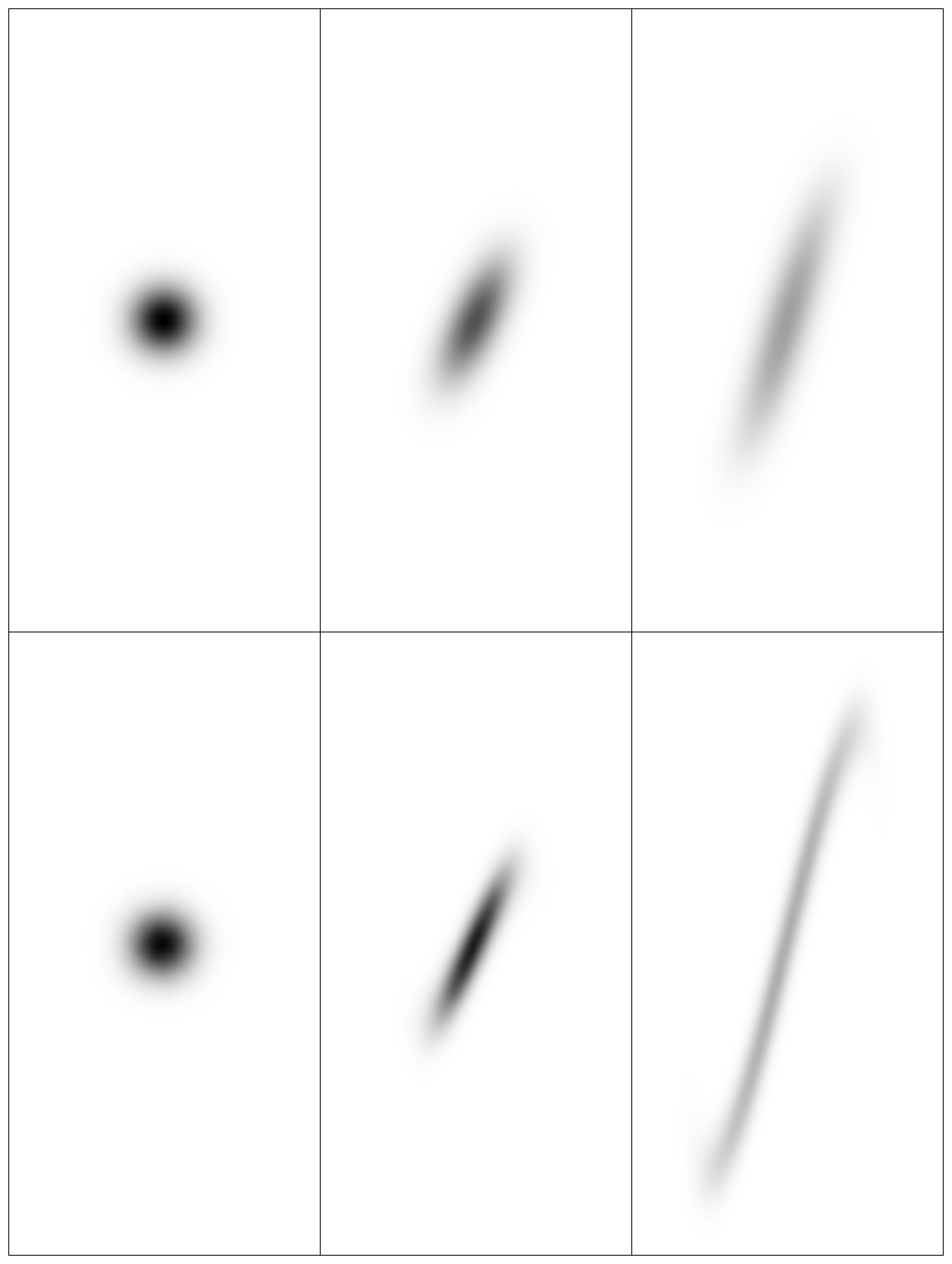}
  \caption{Phase space ($x$,$p$)  evolution of the quantum mechanical Husimi function starting from an instability (top row). Classical Liouville evolution of the same initial function. (bottom row)}
    \label{fig:HusimiAndLiouille}
\end{figure}

In order to illustrate the arguments of the previous section, we show the phase space evolution of the simplest possible system with an instability, a nonrelativistic particle of mass $m$ in the potential 
\begin{equation}
	V(x) = - \alpha x^2 + \beta x^4 
\end{equation}
Around $x=0$, there is an instability in phase space with positive local Lyapunov exponent $\lambda = \sqrt{2 \alpha / m}$. We evolve a minimum uncertainty wave packet centered around $x=0$, $p=0$. Three snapshots of the Husimi function at different instances of time are shown in Fig. \ref{fig:HusimiAndLiouille}, top row. The bottom row shows the classical Liouville time evolution of the same initial functional shape. Evidently, the contraction of the Husimi function in the stable direction is limited compared to the classical evolution. As explained above, this is due to quantum diffusive terms and generically limits the applicability of the classical approximation to the quantum break time.

\end{appendices}

\end{document}